\documentclass[12pt]{article}
\usepackage{graphicx}
 
\usepackage{epsfig}
\usepackage{latexsym}
\usepackage{amsmath}

\begin{document}

\begin{centering}
\title{  God  ($\equiv Elohim$), \\ the first small world network}
 
\vskip0.5cm
\author{ Marcel Ausloos $^{a,b,c}$\footnote{ for correspondence, email address:   
marcel.ausloos@ulg.ac.be  
}  \\
 \\ $^a$
Group of Researchers Applying Physics in Economy and Sociology \\(GRAPES),  Beauvallon, rue de la Belle Jardini\`ere, 483/0021\\
Sart Tilman, B-4031, Li\`ege Angleur, Belgium, Europe  \\
  e-mail: marcel.ausloos@uliege.be
\\$^b$
School of Business, University of Leicester,\\
Brookfield, 
Leicester, LE2 1RQ, UK\\   e-mail: ma683@leicester.ac.uk\\
 $^c$ Department of Statistics and Econometrics,  \\ Bucharest University of Economic Studies,  15-17 Dorobanti Avenue, \\ District 1, 010552, Bucharest, Romania, \\  e-mail: marcel.ausloos@ase.ro
}
\date{\today}
\end{centering}
\maketitle
\vskip -.5truecm
\newpage
\begin{abstract}

In this paper, the  approach of network mapping of words in  literary texts   is extended to ''textual factors'': the network nodes are defined as  ''concepts''; the links are ''community connexions''.  Thereafter, the text network properties  are investigated along  modern statistical physics approaches of networks, thereby relating network topology and algebraic properties, to literary texts contents. As a practical illustration, the first chapter of the Genesis in the Bible is   mapped  into a 10 node network, as in the Kabbalah approach,  mentioning God  ($\equiv Elohim$).   The  characteristics of the network are studied starting from  its adjacency  matrix, and the corresponding Laplacian matrix. Triplets of nodes are   particularly examined in order to emphasize the  ''textual (community) connexions'' of  each   agent "emanation",   through the so called   clustering coefficients   and the  overlap index,  
whence  measuring the ''semantic flow'' between the different nodes.
It is concluded that this  graph is a  small-world network,  weakly dis-assortative, because its average local clustering coefficient  is significantly higher than a random graph constructed on the same vertex set.

\end{abstract}

 


\maketitle

\vskip0.4cm

\section{Introduction   }\label{sec:intro}

 {\it  "Good Lord,   it's a small world, isn't it?"} \cite{MilgramPsyToday}

 An answer is intended here below: 
 
 {\it "Yes, it is: the Good Lord is  a small world network".\\    \qquad ...   \qquad It's even the first one.} \cite{N.B.}

 \vskip0.2cm

In modern statistical physics  \cite{stauffer1}, networks \cite{pastor}, underlying opinion formation of agents located at nodes \cite{1JSM}, with  links defined from data pertaining to social aspects \cite{sociophysics}, have gathered much interest. Many  cases can be found in the literature \cite{costa}. 
Among particularly  interesting  topics, one encounters the case of  finite size networks  in which agents have small connectivity values; such cases are known to be "sociologically more realistic" \cite{MilgramPsyToday,killworth}. 

On the other hand,  texts   carry messages;  they are statistically studied much since Shannon's introduction of the information entropy  definition \cite{shannon}. More recently, it has been discussed that texts can be transformed into  trees \cite{Benedetto,Khmelev} or better into networks in order to study their structure beside  finding word and idea correlations \cite{MasucciRodgers}. 

  Thereafter, one may point to interesting  quantitative considerations about  network related analyses of the characteristics of literary texts; for example, see
\cite{liu2011can} about the morphological complexity of a language,  
\cite{smith2012distinct} about word length frequencies, 
or  about  sequences in Ukrainian texts
  \cite{holovatch2017complex,Buk,Rovenchak2018part}, and still more recently, enjoyable texts analyses  of fables in Slovene as in  \cite{markovivc2019applying,perc2020beauty}. 
  
  
  There are many other papers reporting studies of word and sequences frequencies, or different  language connections as on networks. However to quote all such papers would lead to a  useless digression so far, but see  the recent \cite{markovivc2019applying} which can serve as a recent review, beside these   papers:   \cite{Ling1}-\cite{Ling10}. 
In brief,  the present study pertains to  applications of statistical physics measures and models like those studied in language evolution and in  linguistics 
  \cite{Ling1}-\cite{Ling10}.  
  
In all cases, relevant scientific questions pertain to the dynamics of collective properties, not only of agents on the network, but also  by the network structure itself \cite{lambiotte1}. 
An interesting structure is the "small world network" (SWN),   introduced by Watts and Strogatz  \cite{SWN}. In a SWN,   the neighbors of any given node are both likely to be neighbors of each other and also  be reached from every other node by a small number of linking "steps" \cite{Watts99,Newman00}. I propose to discuss   a 10 node network, as obtained from the first chapter of Genesis \cite{Genesis1}, the so called "Tree of Life",  through the kabbalistic  ($yosher$) tradition \cite{wikisefirot}.

Notice that due to its size,    this Genesis network  might be also expected to become as useful as the karate club  data 
(which has 44 nodes)    \cite{karate} or the acquaintance network of Mormons  
 (which has 43 nodes)    \cite{mormons}, both previously known in the literature for  paradigmatic studies of SWNs\footnote{Other small networks, recently studied,  are  the Intelligent design-Darwin evolution controversy, or financial and geopolitical networks.}. 

 One might wonder why   as  "serious  scientists",  interested in social networks for describing communications between agents, we should  care  about  the  structure  of  such an {\it  a priori}  "mystic network". 
Such  a network is based on information flow between concepts, - not  between words, as it should be emphasized. The matter seems not to have been studied from statistical physics points of view. Nevertheless,  one may sort out \cite{RovenchakBuk2011thermodyn} for a thermodynamic approach.
 Thus, I hope to connect the network analysis methodology  with  that followed in  kabbalistic  studies, - which are much tied to numerology. Moreover,   the present work aims to contribute at introducing a quantitative approach to the analysis  of the interaction between "agents", - here being called sephirot  \cite{Idel1986,Idel2008,Dan2007}, in    small  networks. 

  Thus, in this paper, the previous approaches on text structure studies through word correlations is extended to ''textual factors''. Indeed,  the network nodes are defined as  ''concepts''; the links are ''community connexions''.   The  characteristics of the network are studied starting from  its adjacency  matrix, - its eigenvalues, whence  providing a measure of the ''semantic flows'' between the different nodes. The network Laplacian matrix is also studied along the same lines. 
  Together with Kirchhoff's theorem, and Cheeger's inequality,  the ''spectrum gap''  (between the two smallest eigenvalues) can be used to calculate the number of spanning trees for a given graph  \cite{PREWatanabe}. 
 Indeed, the sparsest cut of a graph can be approximated through the second smallest eigenvalue of its Laplacian by Cheeger's inequality. 
 Furthermore, the spectral decomposition of the Laplacian matrix allows constructing low dimensional embeddings that appear in many machine learning applications and determines a spectral layout in graph drawing, as claimed in $https://en.wikipedia.org/wiki/Laplacian_-matrix$ (accessed on  March 26, 2022).

In    so doing, one adds a quantitative set of values for an answer to a question raised in  \cite{amaralPNAS} on the classes of SWN 
examined in the literature \cite{newmanPNAS01}.   

Besides, the present numerical approach might be in line with modern studies in Kabbalah  research  about numbering \cite{Huss,Garb},
 and quantitative studies on religious adhesion or religiosity  aspects   \cite{religion1}-\cite{HumRep}, as recently used  in socio-physics for examining growth processes, opinion formation, and related topics. This paper is in line with such a frontiers in physics approach.

After introducing. the data set, its origin, in Sect. \ref{sec:dataset}, it seems rather appropriate  to provide  the whole adjacency matrix (10 x 10).  Its construction  goes  in lines with studies on large-scale networks, like co-authorship networks \cite{newmanPNAS01,iina2}.  The present network structural aspects are   first outlined,  before searching for subsequent numerical  and statistical aspects, through a few usual network characteristics in Sect. \ref{sec:stat}. A similar study is performed for the network Laplacian matrix. 

Nevertheless, let it be here pointed out that  triplets of nodes are   particularly examined in order to emphasize the  agent ("emanation") community connexions   through the so called clustering coefficient \cite{SWN} and  the overlap index  \cite{gligorausloos}, in Sect. \ref{sec:clusteringSWN} and Sect. \ref{sec:clusteringAOI}, respectively. The results prove the SWN   nature    of the 10 sephirots  network.
   For completeness,  some other network characteristics, like the assortativity  coefficient \cite{assortativity},  are calculated and reported. 
   A kabbalistic  "generalized point of view" is provided. 


  \begin{figure}
 \includegraphics  [height=15.8cm,width=15.8cm] {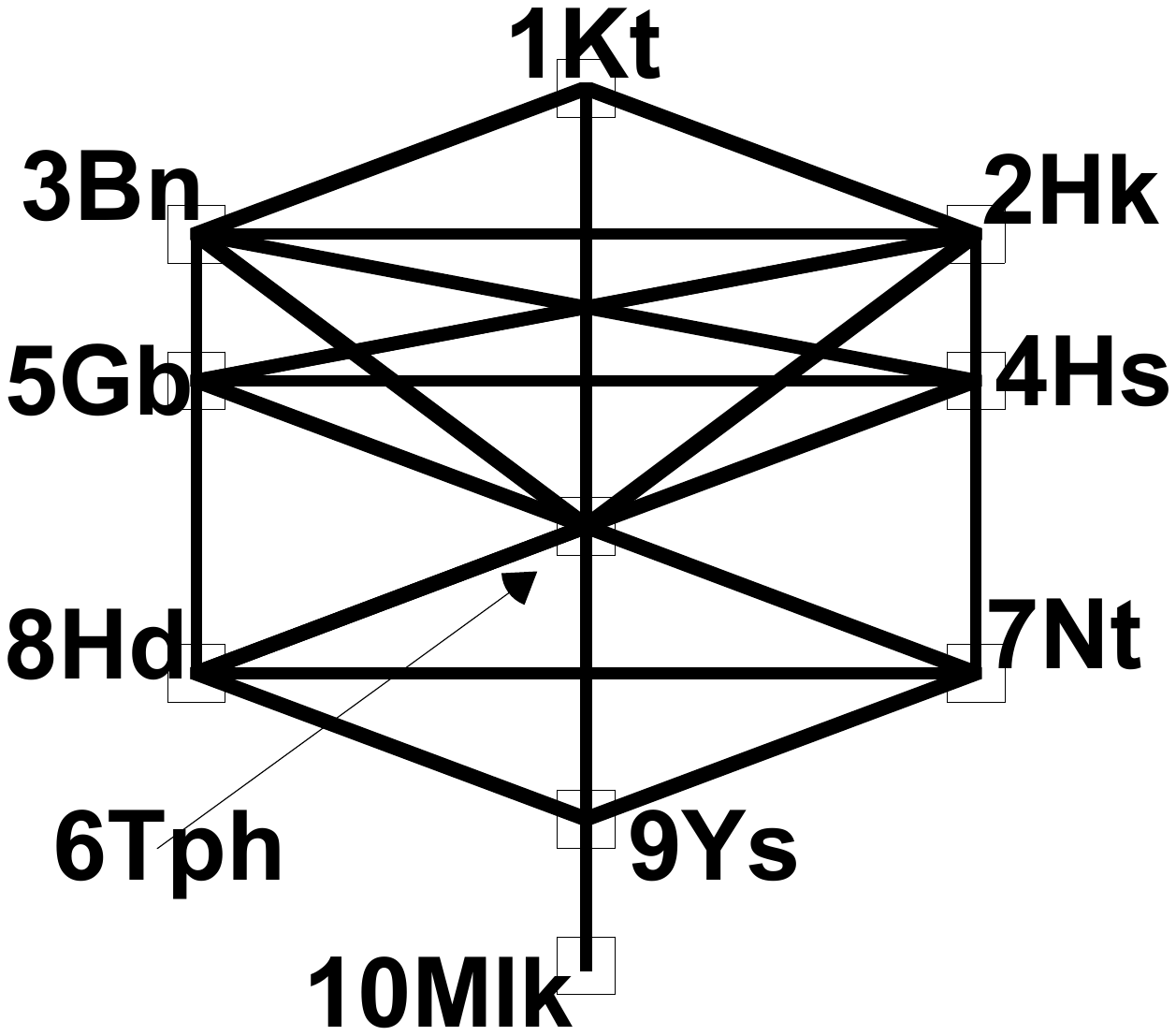} 
 \caption{ The network of 10 sephirots; notations of node labels are found in the main text.}
\label{10godkabal2}
\end{figure}


\section{The data set }\label{sec:dataset}

 Let us consider as  the demonstration of the approach a text in which  concepts are somewhat hidden, - in the present case  within some mystical concept, but without any loss of generality from a theoretical point of view. The data, downloaded from \cite{wikisefirot},  emerges from the kabbalistic interpretation of the occurrence of ''spiritual principles''  at the universe creation. In brief, 
 the Kabbalah\footnote{This paper is not intended to justify of infirm studies of the Bible through Kabbalah methods  \cite{[3],[7]}. 
However, it can be pointed out that the interaction of Kabbalah with modern physics
has generated its own literature, up to
including renaming the elementary particles  with kabbalistic (Hebrew) names or  developing kabbalistic approaches to debates on evolution.}  
 \cite{Idel1986,Idel2008,Dan2007} seems to infer, from the Genesis first chapter, that  "The Infinite" (God) has "emanations" which form a network of ten nodes, like on Fig. \ref{10godkabal2}; the "node  names" are given for further reference in Table \ref{Tnetwork}.
The network so symmetrically displayed  is made of  3 "columns".   (Alternative configurations are given by different schools in the historical development of Kabbalah, with each
articulating different spiritual aspects;  to distinguish the variants is not very relevant  for  the present investigation\footnote{ For example, instead of a  ''tree'' with 3 ''columns'', the  $iggulim$ representation depicts the sefirot as a succession of concentric circles \cite{wikisefirot}. }. The enumeration of the 10 nodes, as on Fig. \ref{10godkabal2},   is stated in the Sefer Yetzirah \cite{Idel1986,Idel2008,Dan2007}.)
 Notice that the Tree of Life  nodes are  arranged onto seven planes; 7 being a mystic  (or sacred) number.
 
 

 

 Between the 10 sephirots, run 22 channels, or paths \cite{comment1}. These   links  are interpreted as the specific connections of 
  ("spiritual")   information flow. In the present case, the flow of information goes according to the node number hierarchy; such a type of directed flow consideration has been recently studied in \cite{Holystentropy2022} in a different context.

In so doing, an adjacency matrix
$G=(g_{ij})\in R^{N \times N}$  can be built,  
with   $ g_{ij}$   = 1  for an existing link  between 2 connected nodes, $\nu_i$ and $\nu_j$, selected among the $N$= 10 nodes here, and $ g_{ij}$ = 0 otherwise, i.e.,

 \begin{equation} \label{eq5.7.G}
   g_{ij} =\left\{ \begin{array}{lcl}
1  \ \   \mbox{\ \ if $\nu_i$ and $\nu_j$  are connected nodes\  }  \\ 
  0  \ \  \ \  \ \  \ \  \ \  \ \   \ \ \  \mbox{\ \ otherwise. \  }   %
\end{array}%
\right.  \end{equation}

Thus, all diagonal terms are $0$;  the matrix is symmetric; it has 44 finite elements, i.e.  $2\;L$, the number of  links. 
   In this  study, the links are neither directional nor weighted;   the nodes have also no "strength".
 
For  further reference, 
let us  here introduce an alternative to the adjacency matrix, i.e. the so called Laplacian matrix of the network:  
$\Lambda =(\lambda_{ij})\in R^{10 \times 10}$, with

 \begin{equation} \label{eq5.8.L}
   \lambda_{ij} =\left\{ \begin{array}{lcl}
-1  \ \  \ \ \ \  \mbox{\ \ if $\nu_i$ and $\nu_j$  are connected nodes\  }  \\ 
d_{\nu_i} \ \  \ \  \ \    \mbox{\ \ if $\nu_i$ $\equiv$ $\nu_j$   \  }  \\ 
  0  \ \  \ \  \ \  \ \   \mbox{\ \ otherwise, \  }   %
\end{array}%
\right.  \end{equation}
where $d_{\nu_i} $  is the degree of the node $\nu_i$, i.e. the number of links at the node. In brief,  the Laplacian matrix $\Lambda $ is the difference between a diagonal matrix $\Delta$  reporting the degree of the node and   the adjacency matrix of the graph.
   For completeness, let us   mention   the finite elements of $\Delta=(d_{ij})\in R^{N \times N}$ through the
 degree list  defined as  $ D = (d_{\nu_1}, d_{\nu_2}, ..., d_{\nu_N})$,
which reads here $D=(3,5, 5, 5, 5, 8, 4, 4, 4, 1)$

Thus, the adjacency matrix reads 
\begin{eqnarray}
G=\left( \begin{tabular}{llll  llll  ll} 
-&1&1&-&-&1&-&-&-&-\\ 
1&-&1&1&1&1&-&-&-&-\\ 
1&1&-&1&1&1&-&-&-&-\\ 
-&1&1&-&1&1&1&-&-&-\\ 
-&1&1&1&-&1&-&1&-&-\\ 
1&1&1&1&1&-&1&1&1&-\\ 
-&-&-&1&-&1&-&1&1&-\\ 
-&-&-&-&1&1&1&-&1&-\\ 
-&-&-&-&-&1&1&1&-&1\\ 
-&-&-&-&-&-&-&-&1&-\\ 
\end{tabular} \right)    \label{G10x10}
\end{eqnarray} 

in which  each 0  is replaced by a - for better readability.
The $\Lambda$ matrix is written    and analyzed below. 

\begin{table}
\begin{tabular}{|c|c||c|c|c||c|c|c|c|c|c|c|c|c|} 
 \hline $G$  & 	 node 	&	n.links &	$p_i$    & 	 $q_i$   &	      &   n.triads  & 	VCC  & 	LCC   & 	\\ 		
 matrix & 	 name  	&	 $d_{\nu_i}$ &	$(\%)$    & 	$(\%)^{2}$  &	$d_{\nu_i} (d_{\nu_{i}-1})/2$        &   	$e_i$ & 	    $c_i$ &     $ \Gamma_i$  & 	\\ \hline	\hline							
1Kt  & 	Kether	 & 	3 & 	6.82 & 	9.21 & 	3 & 	3 &	1 & 	6/6 & 	\\ \hline							
2Hk & 	Hokm.	 & 	5 & 	11.4 & 	25.6 & 	10 & 	 6  &	0.6  & 	13/15 &	\\ \hline							
3Bn & 	Binah	 & 	5 & 	11.4 & 	25.6 & 	10 & 	 6 & 	0.6 & 	13/15 & 	\\ \hline							
4Hs & 	Hesed	 &  	5 & 	11.4 & 	25.6 & 	10 & 	7 & 	0.7 & 	12/15 & 	\\ \hline							
5Gb & 	Geb.	 &  	5 & 	11.4 & 	25.6 & 	10 & 	7 & 	0.7 & 	12/15 & 	\\ \hline							
6Tph & 	Tiph.	 &   	8 & 	18.2 & 	65.5 & 	28 & 	13 &	0.464 & 	21/36 & 	\\ \hline							
7Nt & 	Netsah	 &  	4 & 	9.09 & 	16.4 & 	6 &	3 & 	0.5 & 	8/10 &	\\ \hline							
8Hd & 	Hod	&	4 & 	9.09 & 	16.4 & 	 6 & 	3 & 	0.5 & 	 8/10  & 	\\ \hline							
9Ys & 	Yesod	 &  	4 & 	9.09 & 	16.4 & 	6 &	3 & 	0.5 & 	7/10 & 	\\ \hline							
10Mlk   &	Malk.	 &  	1 &	2.27 &	1.02 &	 0 &	0       &	0   &	1       &	\\ \hline																 
\end{tabular} 
 \begin{centering}
 \caption{ Characteristics of the network   matrix $G$, with  hereby defined node ($i$) notations, (Hokm. =  Hokmah; Geb. = Gebourah;  Tiph. = Tiphereth; Malk. = Malkouth) in the first and second  columns, and their corresponding number of links ($d_{\nu_i}$). The values of usual structural information  for networks are given: the probability    $p_i$   that a  vertex $\nu_i$  has  a degree  $ d_{\nu_i}$; $q_i$ is fully defined in Eq.(\ref{q}) in terms of $p_i$   and
  the $i$ vertex  degree  $ d_{\nu_i}$;  
  the  possible maximum number of   different wedges, ($d_{\nu_i} (d_{\nu_i}-1)/2$); the number of  triads ($e_i$) associated to a given node ($i$)  in $G$;  VCC, the corresponding  clustering coefficient ($c_i$) of a vertex $i$, - from which one deduces the global clustering coefficient (GCC) {\it of the network}, and
 $ \Gamma_i$,  the local clustering coefficient (LCC) of a vertex $i$, -   from which one deduces the  average local clustering coefficient   {\it for the network}, see Sect.   \ref{sec:clusteringLCC}. }\label{Tnetwork} \end{centering}  
 \end{table}

\begin{table} \begin{centering} \begin{tabular}{c|llll  llll  llll} 
$N_{i,j}$&1&2&3&4&5&6&7&8&9&10\\\hline
1&-&2&2&-&-&2&-&-&-&-\\ 
2&2&-&4&3&3&4&-&-&-&-\\ 
3&2&4&-&3&3&4&-&-&-&-\\ 
4&-&3&3&-&3&4&1&-&-&-\\ 
5&-&3&3&3&-&4&-&1&-&-\\ 
6& 2&4&4&4&4&-&3&3&2&-\\ 
7&-&-&-&1&-&3&-&2&2&-\\ 
8&-&-&-&-&1&3&2&-&2&-\\ 
9&-&-&-&-&-&2&2&2&-&-\\ 
10&-&-&-&-&-&-&-&-&-&- \\\hline
\end{tabular} 
\caption{ The ($N_{i,j}$) number of  different  (undirected information flow) paths  between two nearest neighbors $i$ and $j$,  through a nearest neighbor $k$.}
  \label{TlinkstatNij} 
 \end{centering} \end{table}
 
Anyone knows that when there is   a matrix,  one   looks for eigenvalues and eigenvectors :
the (necessarily real) eigenvalues are found to be  equal to:



5.02314,
2.21045,
0.61803,
0.13191,
0.00000,
-1.00000,
-1.36550,
-1.61803,
-2.00000,
-2.00000.

They are distributed in a (quasi logarithmically) decreasing order:
$y = 4.503 - 6.865$\;$log(x)$,   with $R^2$= 0.977.

Thereafter,    one   can look  for the 10 eigenvectors; however, they are not shown  for saving space, -  their writing being irrelevant within the present aim. Nevertheless, the above suggests that  a Principal Component Analysis can be a  complementary valuable investigation for "community detection".

 The network Laplacian matrix reads
  \bigskip
 \begin{eqnarray}
\Lambda=\left( \begin{tabular}{llll  llll  ll} 
3& -1 & -1 &-&-& -1 &-&-&-&-\\ 
 -1 &5& -1 & -1 & -1 & -1 &-&-&-&-\\ 
 -1 & -1 &5& -1 & -1 & -1 &-&-&-&-\\ 
-& -1 & -1 &5& -1 & -1 & -1 &-&-&-\\ 
-& -1 & -1 & -1 &5& -1 &-& -1 &-&-\\ 
 -1 & -1 & -1 & -1 & -1 &8& -1 & -1 & -1 &-\\ 
-&-&-& -1 &-& -1 &4& -1 & -1 &-\\ 
-&-&-&-& -1 & -1 & -1 &4& -1 &-\\ 
-&-&-&-&-& -1 & -1 & -1 &4& -1 \\ 
-&-&-&-&-&-&-&-& -1 &1\\ 
\end{tabular} \right)    \label{Lambda10x10}
\end{eqnarray}
 The eigenvalues are:

9.01939,
6.61803,  
6.48072,
6.00000,  
5.13659,  
4.38197,  
3.48940,
2.13004,
0.74387,
0.00000.
 
They are distributed according to: $ y \simeq 9.478 - 0.923 \; x$; with   $R^2$= 0.971.
 
  Together with Kirchhoff's theorem, the Laplacian matrix eigenvalue spectrum can be used to calculate the number of spanning trees for a given graph, $\eta$. 
 The sparsest cut of a graph can be approximated through the second smallest eigenvalue of its Laplacian, i.e. $\lambda_2 = 0.74387$, here,  by Cheeger's inequality.  Since the Laplacian matrix spectral gap is also obviously equal to 0.74387,  one finds
 \begin{equation} \label{numberoftrees}
 \frac{\lambda_2}{2} \; \leq  \eta \;   \sqrt{\lambda_2 ( 2\;  d_{\nu_j}^{(M)} -\lambda_2)}
 \end{equation} 
 where  $ d_{\nu_j}^{(M)} (\;=8)$ is   the largest node degree.  Thus, $0.372\leq \eta \leq 3.369$.
 



\section{Data  statistical analysis } \label{sec:stat}

Next, one  proceeds performing some classical structural analysis as usual  on such networks, i.e. an analysis of indicative coefficients: one obtains the network node in- and out-degree distributions,  the network assortativity, the (Global and Local) Clustering Coefficients,  and  the Average Overlap Index. 

In the present case, the matrix, or network, is symmetric, whence   the number of links exiting from a node $\nu_j$, i.e. the $out-degree$, is equal to the $in-degree$ number.  The largest degree (=8) is for node 6; the smallest (=1) is for node 10; the average degree, counting both  $out-degrees$ and  $in-degrees$ is easily found to be 4.4.

On Table \ref{Tnetwork}, one  also gives for each node, the  degree, i.e. the number of links  ($d_{\nu_i}$) exiting from  or  entering into each node  ($i$). 
 Table  \ref{Tnetwork}  also reports the  possible maximum number of   different wedges, ($d_{\nu_i} (d_{\nu_i}-1)/2$), and triads,  ($e_i$), associated to a given node ($i$)  in $G$.

 In addition,  we  report he  number    ($N_{i,j}$)  of different  paths going  through a nearest neighbor $k$ of two nearest neighbors $i$ and $j$ in Table \ref{TlinkstatNij}. This number is equivalent to the number of triangles sharing the link $(i,j$).

\section{Clustering }\label{sec:clusteringSWN} 

The tendency of the network nodes to form local interconnected groups is a convincing arguments for describing social networks along the statistical physics modern formalism. Such a behavior is usually quantified by a measure referred to as the clustering coefficient \cite{SWN}.
The amount of studies on this characteristic of networks has led to the particularization of the term in order to focus on different complex features of   networks. Here, one  considers the global clustering coefficient and the local clustering coefficient, together with the overlapping index, and the assortativity for a text mapped into a network.

  Indeed, the most relevant elements of a heterogeneous agent interaction network can be identified by analyzing    global and  local connectivity properties.       In the present case, this    can be attempted by analyzing the number of triangles with   agent (or "emanation") nodes belonging to the same "community"  or not,  depending on the type of 
  connexions.  The former number gives some hierarchy information; the latter some reciprocity measure, i.e. recognition of leadership or proof  of some challenging conflict among the emanations.

 \subsection{Global Clustering Coefficient}   \label{sec:clusteringGCC}
 
The global clustering coefficient (GCC) {\it of the network} is defined as   $<c_i>$, the average of $c_i$ {\it over all the vertices} in the network, $<c_i>\;=\sum c_i /N$, where $N$ is the number of nodes of the network, and where the clustering coefficient $c_i$ of a vertex $i$ is given by the ratio between the number $e_i$ of triangles  sharing that  specific vertex $i$, and the  maximum number of triangles that the vertex could have. If a node $i$ has $d_{\nu_i}$  neighbors, then the so called  clique, i.e.  a complete graph in fact,  would have $d_{\nu_i} (d_{\nu_i}-1)/2 $ triangles at most, thus one has,
\begin{equation}
  c_i \;   = \; \frac{2}{d_{\nu_i} (d_{\nu_i}-1)} \; e_i 
 \end{equation}  \label{GCC}. 

The value of GCC is found to be $<c_i>\;=\;0.5564$, from the raw data in Table \ref{Tnetwork}.
   
  \subsection{Local Clustering Coefficient} \label{sec:clusteringLCC}
In the literature  \cite{newmanPNAS01},   the term 'clustering coefficient' refers to various quantities, relevant to understand the way in which nodes   form communities, under some criterion.
By definition, the ''local clustering coefficient'' (LCC) $\Gamma_i$ for a node $i$ is the number of  
 links between the vertices within the nearest neighbourhood of $i$ divided by the maximum number of links that could possibly exist between them.
  It is relevant to note that the above  GCC is not trivially related to the LCC, e.g. the GCC is not the mean of LCC.  In the former case,   triangles having common edges are emphasized, in the latter case only the number of  links is relevant. This number of links common to triangles sharing the node $i$  can vary much with the number of connected nearest neighbour nodes indeed.  Basically,  the GCC value quantifies how much the neighbors of $i$   are close to being part of a   complete graph. In contrast, LCC rather serves to determine whether a network is a SWN \cite{SWN} or not.

The LCC  ($\Gamma_i$) values are    given in Table \ref{Tnetwork}, under a  ratio form in order to emphasize that the numerator of the fraction is the sum of $d_{\nu_i}$ plus  the number of links making triangles in the nearest neighborhood of $i$, while the denominator is obviously  $d_{\nu_i} (d_{\nu_i}+1)/2$.      It is easily deduced that $<\Gamma_i>=0.8217$. 

There is no drastic conclusion  to draw from this specific value, since not many corresponding values are reported in the literature allowing a comparison with other networks \cite{footnote3}.
Yet, let it be recalled   that the lower the $\Gamma_i$ values,  the less "fully connected" appears to be the network. This is not the present case.

However,  let it be emphasized that  a  graph is considered to  be {\it small-world}, if its average local clustering coefficient  is significantly  higher than a random graph constructed on the same vertex set, i.e. here with $N=10$.  Thus, one confirms  that the present network   looks like a SWN rather than either a random network (RN)  
or a complete graph (CG).   

\section{Average Overlap Index}  \label{sec:clusteringAOI}

Finally, for characterizing members of communities, in another hierarchical way,  let us also calculate the Average Overlap Index (AOI) $O_{ij}$; its mathematical formulation and its properties are found in \cite{gligorausloos}
in the case of a unweighted network made of $N$  nodes linked by
 $(ij)$ edges,

\begin{equation}  \label{o} 
O_{ij} = \frac{{N_{ij} \left( {d_{\nu_i }+ d_{\nu_j} }
\right)}}{{4\left( {N- 1} \right)\left({N - 2} \right)}},\quad
\quad i \ne j
\end{equation}
where $N_{ij}$ is the measure of the \underline{common number} of  (connected) neighbors to the
\textit{i}
and \textit{j} nodes. 
In the present case:$ 4 ( N- 1)(N - 2) =288$
N.B. in a fully connected network, $N_{ij} = N - 2$. Of course, $O_{ii}=0$  by definition.

The \textit{Average Overlap Index} for the node   $i$ is
defined as:
\begin{equation}  \label{average}
\left\langle {O_i} \right\rangle = \frac{1}{{N - 1}}\; \sum\limits_{j= 1}^N {%
O_{ij}. }
\end{equation}

The  values are given in Table 3. 

\begin{table} \begin{centering}   \begin{tabular}{clcllll  llll  llllcll} 
[288\; $O_{i,j}$]&1&2&3&4&5&6&7&8&9&10&&$<O_i>$\\\hline
1&	   0 &    16 &    16 &     0 &     0 &    22 &     0 &     0 &     0 &     0 & 	&0.01230&\\
2&	   16 &     0 &    40 &    30 &    30 &    52 &     0 &     0 &     0 &     0 & 	&0.03825\\
3&	   16 &    40 &     0 &    30 &    30 &    52 &     0 &     0 &     0 &     0 & &0.03825&\\
4&	    0 &    30 &    30 &     0 &    30 &    52 &     9 &     0 &     0 &     0 & 	&0.03438&\\
5&	    0 &    30 &    30 &    30 &     0 &    52 &     0 &     9 &     0 &     0 & 	&0.03438&\\
6&	   22 &    52 &    52 &    52 &    52 &     0 &    36 &    36 &    24 &     0 & 	&0.07423&\\
7&	    0 &     0 &     0 &     9 &     0 &    36 &     0 &    16 &    16 &     0 & 	&0.01753&\\
8&	    0 &     0 &     0 &     0 &     9 &    36 &    16 &     0 &    16 &     0 & 	&0.01753&\\
9&	    0 &     0 &     0 &     0 &     0 &    24 &    16 &    16 &     0 &     0 & 	&0.01275&\\
10&	    0 &     0 &     0 &     0 &     0 &     0 &     0 &     0 &     0 &     0 & 	&0&\\
\hline
$\sum_j$&54 &168 &168 &151 &151 &326 &77 &77 &56 &0 &&\\
\end{tabular}   \label{Oij} 
\caption{[288\;$O_{i,j}$]: the numerator of the overlap index, Eq.(\ref{o}),  of the neighboring  $\nu_i$ and $\nu_j$ nodes; and $<O_i>$, the average overlap index, Eq. (\ref{average}). 
\\N.B.  
$ 288= 4 ( N- 1)(N - 2)$,  while $\sum_i \sum_j O_{i,j}$ =1228.}
 \end{centering}   \end{table}

 This measure, $\left\langle {O_i} \right\rangle$, can be interpreted indeed as an other  form of clustering attachment  measure: the higher the number of nearest neighbors, the higher the $\left\langle {O_i} \right\rangle$, the more so if the $i$ node has a high degree $d_{\nu_i}$. Since the summation is made over  all possible $j$ sites connected to $i$ (over  all  sites in a fully connected graph), $\left\langle {O_i} \right\rangle$  expresses  a measure of the local (node) density near the $i$ site.

Recall also that  in   magnetic networks,   the
links are like exchange integrals between spins located at $i$ and $j$.
An average over the exchange integrals provides an estimate  of the critical (Curie) temperature at which a spin system undergoes an order-disorder transition, - and conversely. Therefore $\left\langle {O_i} \right\rangle$  can also  be interpreted, in a physics sense,  as a measure of the stability of the node versus perturbations due to  an external (for example, thermal) cause.  In other words, in the present context, a high $\left\langle {O_i} \right\rangle$  value reflects the $i$ node strong attachment to its community: the main ''textual factor'' is thereby emphasized.  Here, for our illustrative example, the highest value ($\simeq 0.074$) correspond to the  6th node (''emanation''),  Tiphereth, - as should have been also visually expected from Fig. 1.

The \textit{average overlap index} of each node, obtained according to Eq.(\ref{average}), 
  are given in  Table 3. 
The order of magnitude of the  $\left\langle             {O_i} \right\rangle$ values are $\sim  0.05 $,  much smaller  than in other investigated cases, like in  \cite{gligorausloos} (or \cite{AOILA}). This is due to the low value of $N_{ij}$, in the present case. 

For completeness, observe that  $\sum_i \sum_i O_{i,j}$ =1228, whence 1288/288 = 4.472 , 
 which divided by $N$ leads to $\sim  0.4472$, as another characteristic of the  average overlap number of triangles throughout the network. 

In order to indicate some aspect of the attachment process in a network, one can calculate its so called "assortativity"  \cite{assortativity}. The term refers to a preference for a network node to be attached to others,  depending on one out of many node properties \cite{assortativity}.  Assortativity is most often measured after a (Pearson) node degree correlation coefficient $r$

\begin{equation}\label{r}
r=\frac{\sum_{i,j=1}^N q_i q_j g_{ij}- [\sum_{i,j=1}^N (q_i+q_j) g_{ij}]^2 /L}{ \sum_{i,j=1}^N( (q_i^2 +q_j^2)g_{ij}) -[\sum_{i,j=1}^N (q_i+q_j)g_{ij}]^2/L }
\end{equation}
where  
\begin{equation}\label{q}
  q_i=\frac{k^{}_i p^{}_{i}}{\sum_i k^{}_i p_i},
  \end{equation}
  where $ k_i$ is  the $i$ vertex (total) degree  $ d_{\nu_i}$, 
in which  $p_i$   is  the probability that a  vertex $i$  has  a degree  $ d_{\nu_i}$  (this can be here obtained/read from Fig. \ref{10godkabal2} or Table  \ref{Tnetwork}):
 $ L$  is the number of
connecting channels (= 22, here);
$r=1$ indicates perfect  assortativity, $r=-1$ indicates perfect "dis-assortativity", i.e. a perfectly negative correlation.
 
 For the  (text based) network of interest here, we have found,  $r= -0.229$, a quite negative value  for the $assortativity$ notion, in most networks. 
The present finding  is somewhat surprising, since according to   \cite{assortativity},  almost all ''non-social networks''  \cite{assortativity}   seem to be quite dis-assortative, even though  the ''social networks''   usually  present significantly assortative mixing. However,  the technological and biological networks  usually  are all dis-assortative:   the hubs are (primarily) connected to less connected nodes,  $dixit$ Newman \cite{assortativity}. The present case is a  weakly  dis-assortative network.
 
In order  to show a positive value of $r$, a network must have some specific additional structure that favors assortative mixing, i.e. a division into communities or groups; {\it a contrario},  to see significant dis-assortativity, the highest degree vertices in the network need to have degree on the order of $\sqrt{N}$, where $N$ is the total number of vertices, so that there is a substantial probability of some vertex pairs sharing two or more edges. Here  $\sqrt{22} \simeq 4.69$; the highest degree which is for $Tiphereth $ = 8 is (at once visually found from Fig. \ref{10godkabal2}) the "knowledge transfer hub", - the most important emanation.
 
 One may  consider the practical aspects resulting from the node characteristics, next those from links.
In relation to a "generalized  kabbalistic point of view", one may make the following comments.

Let us observe two new integer numbers appearing through the study:
288\footnote{ It is thought that the earth’s average surface temperature = 288 K, but that might neither be relevant, nor suggests further investigation.}
 and 1228.  Notice that
\begin{itemize}
\item
288: this number contains profound significance; in Kabbalah, it
refers to the number of “sparks” that God had to remove in
order to create the world; see $https://www.biblegematria.com/288-holy-sparks.html$ (Accessed March 01, 2022)
\item
1228: the Hebrew name of Simon Peter, Symehon  Hacephi, is 1228 in Hebrew name numeration; (\cite{magician}, p. 54).
\end{itemize}

Comments and suggestions on  such  a "society structure" within formal texts can be  thought to arise from similar numerical perspectives. 
 
 \section{Conclusions}
\label{sec:conclusions}

  In frontiers science, 
  prior to scientific excitations and paper avalanches,   there are modest inter-connections, between  authors and between fields.
This is one of the underlying ideas for the  present problem, not at the level of authors but at the semantic level, - justifying the study.  Two apparently unrelated research fields are interconnected.  One can study texts through network mappings, - nothing new. 
 I recall that the Ukrainian language network used in the selected fables  studied in \cite{holovatch2017complex} is a strongly correlated, scale-free, small world network. 
In the present case, one goes a little bit further: instead of another word correlation study, one examines  textual concept distributions. Moreover, picking up a basic text with some mystic ingredient, one covers a wide gap between various disciplines,   with a physics support.

One has proposed to examine a theoretical question on applied linguistics, with a specific illustration, but in so doing also asked : do the sephirot, thus  nodes and  links of a mystic network,  mean something from a statistical physics point of view, knowing their "esteem" or "sense"  in kabbalistic work?   Thus, in fact, the study has some similarity  to other "social network" considerations: {\it mutatis mutandis},  in the present problem  the agents are the sephirot, while  the links carry the information flows between emanations. 

Practically,  the $yosher$  kabbalistic mapping of a selection of concepts from the Genesis in the Bible produces a network \cite{wikisefirot}. In order to characterize the  necessarily small network, based on  its adjacency matrix, one  has  calculated  a few {\em specialisation coefficients}.
Surely, in future work, one could consider many other quantities of interest for networks \cite{bernard}; the matter is left for the imagination of concerned researchers.

In particular, assortativity  characteristics of the network have been examined,   - in so doing searching whether there is a proof of  any preference of a sephirot  attachment to  some sub-networks.   Examining the whole network, through their communities and the inter-community links,  it is found that  the sephirots are neither perfectly assortative nor perfectly  dis-assortative.
From the  values of the Pearson node degree correlation coefficient $r$  it is asserted that the network is rather  dis-assortative,  - but weakly correlated in contrast to the fables in \cite{holovatch2017complex}.
 This is contrasted to fictional social networks  \cite{PRE68.-3NewmanPark,ChoiKim2007directedmytho} which are found to  be small-world, highly clustered, and hierarchical,  which typically differ from real ones in connectivity and levels of assortativity  
\cite{markovivc2019applying}.

According to \cite{PRE68.-3NewmanPark},  a clustering coefficient  can (also) be defined by
 \begin{equation}
 C= \frac{1}{N} \;\frac{[<\kappa^2>- < \kappa  >]^2}{<\kappa>^3},
 \label{clusteringNP} \end{equation}
 where $\kappa$  is the excess degree; in the present text case,  $<\kappa >$ = 3.4.
 Thus, $C=0.308$.

In order to characterize in greater detail the intercommunity  structure  complexity, - its ''information flow'',   one  can also consider elementary entities made of a few sephirots. The smallest  (geometric) cluster to be examined is the triangle. 
  In this respect the study of the  local clustering  coefficients  indicates  a low value for these inter-community sub-networks. The average overlap index (AOI) \cite{gligorausloos}  allows to extract  from the  clusters those nodes which  inside their community and with respect to the others are the centers of more attention. One may claim that one gives some scientific (statistical physics) emphasis  to one kabbalistic emanation.

 From a fundamental statistical physics point of view, one may emphasize the "added value" of the present investigation.  Return to Amaral et al.  \cite{amaralPNAS} who have proposed three classes of SWN: (i) scale-free networks, characterized by a vertex connectivity distribution that decays as a power law; (ii) broad-scale networks, characterized by a connectivity distribution that has a power law regime followed by a sharp cutoff; and (iii) single-scale networks, characterized by a connectivity distribution with a fast decaying tail. The   analyses presented in the main text suggest that the  network belongs to the third  category. It should be of course of interest to find out if this conclusion holds  for other  ''textual factors'' in other literary texts.

 
Finally, let it be recalled that some time ago,   {\it "God is a mathematician"}, was concluded by Newman \cite{KNewman} and questioned by  Livio \cite{MLivio03}.
Elsewhere, one may find the question: {\it "Is God a geometer?" }\cite{StewartGolubitsky2010}.

Apparently, according to the present text analysis of the Genesis, God  ($\equiv Elohim$) is also the (chronologically) first small world network, - for the monotheistic religions.

\bigskip 

{\bf Acknowledgements:}
 Thanks to Prof. Emmanuel E. Haven and Prof. Claudiu
Herteliu, for numerous and encouraging comments.
Thanks also to Prof. Roy Cerqueti and Prof. Krzysztof
Kulakowski for comments on a previous version.
Moreover, special thanks go to Rabbi Mark L. Solomon for
his very kind highlighting comments.

\bigskip 

{\bf Conflict of interest  statement:}
The author declares that the research was conducted in the absence of any philosophical, commercial or financial
relationships that could be construed as a potential conflict of interest.

\bigskip 

{\bf Funding: } Work partially supported by the 
Romanian National Authority for Scientific Research and Innovation, under UEFISCDI  
PN-III-P4-ID-PCCF-2016-0084
research grant.

\bigskip 

{\bf Ethics:} no animal, no human was mistreated during this study, to the best knowledge of the author. 
\begin{itemize}
\item 
{\bf Studies involving animal subjects.
Generated Statement: } No animal studies are presented in this manuscript.
\item
{\bf Studies involving human subjects.
Generated Statement: }No human studies are presented in this manuscript.
\item
{\bf Inclusion of identifiable human data.
Generated Statement: }No potentially identifiable human images or data is presented in this study.
\end{itemize}


\begin{thebibliography}{99}
\bibitem{comment1} Such a number of links can be associated with the 22 letters of the Hebrew alphabet
\bibitem{footnote3} The clustering    coefficient    of    a    random    graph    is  equal to $<k>/N$, i.e.,  = 0.44 in the present text case.
\end{thebibliography}
 \end{document}